\documentstyle[12pt,epsf]{article}

\def\tr{\mathop{\rm tr}\nolimits}
\def\Tr{\mathop{\rm Tr}\nolimits}

\renewcommand{\theequation}{\thesection.\arabic{equation}}
\makeatletter
\@addtoreset{equation}{section}

\newcommand{\VEV}[1]{\left\langle #1 \right\rangle}
\newcommand{\del}{\partial}
\newcommand{\Mp}{M_P}

\begin{document}
\baselineskip 0.2in

\begin{titlepage}

\begin{flushright}
YITP-99-5\\
UT-836\\
KUNS-1559\\
TUM-HEP-344/99\\
SFB-375-333
\end{flushright}

\vspace{4ex}

\begin{center}
{\large \bf
Anomalous U(1) Gauge Symmetry and \\
Lepton Flavor Violation}

\vspace{6ex}

\renewcommand{\thefootnote}{\alph{footnote}}

Kiichi Kurosawa\footnote
{
e-mail: kurosawa@hep-th.phys.s.u-tokyo.ac.jp
}
and
Nobuhiro Maekawa\footnote
{
He is visiting TU from Oct.98 to Sep. 99. 
e-mail: maekawa@physik.tu-muenchen.de
}

\vspace{4ex}
${}^{a}${\it YITP, Kyoto University, Kyoto 606-8502, Japan}\\
${}^{a}${\it Department of Physics, University of Tokyo,\\
      Tokyo 113-0033, Japan}\\
${}^{b}${\it Department of Physics, Kyoto University,\\
     Kyoto 606-8502, Japan}\\
${}^{b}${\it Physik Department, Technische Universit\"at M\"unchen,\\
     D-85748 Garching, Germany}
\end{center}

\renewcommand{\thefootnote}{\arabic{footnote}}
\setcounter{footnote}{0}
\vspace{6ex}

\begin{abstract}

In recent years, many people have studied the possibility that the
anomalous $U(1)$ gauge symmetry is a trigger of SUSY breaking and/or an
origin of the fermion mass hierarchy. Though it is interesting that the
anomalous $U(1)$ symmetry may explain these two phenomena
simultaneously, it causes a negative stop mass squared or a severe
fine-tuning in order to avoid the FCNC problem. Recently, it was pointed
out that the $F$-term contribution of the dilaton field can dominate
the flavor-dependent contribution from the anomalous $U(1)$ $D$-term,
so that the FCNC problem may be naturally avoided. In this paper, we 
study the case in which the dilaton is
stabilized by the deformation of the K\"ahler potential for the dilaton
and find that the order of the ratio of the $F$-term to the
$D$-term contributions is generally determined. This implies that the
branching ratio of $\mu\to e\gamma$ can be found around the present
experimental bound. 
\end{abstract}

\end{titlepage}

\section{Introduction}
One of the most important problems of the standard model is the stability 
of the weak scale and another is the hierarchy problem of fermion
masses. The most promising solution for the former problem is to
introduce supersymmetry (SUSY). If SUSY is realized in nature, we should
understand the origin of SUSY breaking. One of the possibilities
is that an anomalous $U(1)$ gauge symmetry with anomaly cancellation due
to the Green-Schwarz mechanism \cite{GS} triggers SUSY breaking
\cite{BD,DP} and mediates the SUSY breaking effects in ordinary
matter \cite{mediate}. On the other hand, it is widely known that the
latter problem of the fermion mass hierarchy can be solved by assignment of
anomalous $U(1)$ charges to the matter fields \cite{FN,yukawa}. It
is interesting to examine the possibility that the anomalous $U(1)$
gauge symmetry explains both the origin of SUSY breaking and
the origin of fermion mass hierarchy \cite{NW,Zhang}. However, there are 
some unsettled issues for this scenario. In order to explain the fermion mass
hierarchy, we should assign anomalous $U(1)$ charges dependent on
the flavor. This induces  non-degenerate scalar fermion masses
through the anomalous $U(1)$ $D$-term. The large SUSY breaking scale allows us 
to avoid the flavor changing neutral current (FCNC) problem \cite{DG,CKN},
while it causes a negative stop mass squared or a severe fine-tuning
\cite{neg_stop}.

Recently, Arkani-Hamed, Dine and Martin \cite{ADM} pointed out that the
$F$-term contribution of the dilaton field can be larger than the
anomalous $U(1)$ $D$-term contribution, depending on how the dilaton is
stabilized. If the $F$-term contribution dominates the $D$-term
contribution, the situation is drastically changed. Through loop
corrections of the gaugino, the $F$-term contribution can induce 
degenerate scalar fermion masses. As a result, the constraints from FCNC
processes are weakened.

In this paper, we point out that the order of the ratio of the
$D$-term to the $F$-term contributions is generally determined when the
dilaton is stabilized by smallness of the second derivative of the
K\"ahler potential for the dilaton. The $D$-term gives the
flavor-dependent sfermion masses, while the $F$-term of the dilaton
contributes the flavor-independent sfermion masses through loop
corrections of the gaugino. Therefore FCNC processes can be predicted by
this ratio. In this scenario the most dangerous process is the
$\mu\to e \gamma$ process, and it can be found around the present
experimental bound to the branching ratio $BR(\mu\to e\gamma)<4.9\times
10^{-11}$ \cite{PDG}. We also investigate the case in which there are other
contributions to sfermion masses as well as loop corrections of the
gaugino. Even in this case, the branching ratio is not so different from
the current bound.

\section{Anomalous $U(1)$ gauge symmetry}
First we review the anomalous $U(1)_A$ gauge symmetry. It is
well-known that some low energy effective theories of string theory
include the anomalous $U(1)_A$ gauge symmetry that has non-zero
anomalies, such as the pure $U(1)_A^3$ anomaly, mixed anomalies with the
other gauge groups $G_a$, and the mixed gravitational anomaly
\cite{U(1)}.\footnote
{
For example, conditions for the appearance of anomalous $U(1)$ in
orbifold string models are discussed in Ref.~\cite{KN}.
} 
These anomalies are canceled by combining the nonlinear transformation
of the dilaton chiral supermultiplet $S$ with the gauge transformation
of the $U(1)_A$ vector supermultiplet $V_A$:
\begin{eqnarray}
V_A &\to&  V_A
  +\frac{i}{2}\left(\Lambda-\Lambda^\dagger\right),\\
S &\to& S+\frac{i}{2}\,\delta_{GS} \,\Lambda,
\label{trans}
\end{eqnarray}
where $\Lambda$ is a parameter chiral superfield. This cancellation occurs
because the gauge kinetic functions for $V_A$ and the other vector 
supermultiplets $V_a$ are given by
\begin{equation}
{\cal L}_{\mbox{\scriptsize gauge}}
 = \frac{1}{4} \int d^2\theta \left[\,k_A S \,W_A{}^\alpha
W_A{}_\alpha 
       + k_a S\, W_a{}^\alpha W_a{}_\alpha \,\right] + \mbox{h.c.}, 
\end{equation}
where $W_A{}^\alpha$ and $W_a{}^\alpha$ are the chiral projected
superfields from $V_A$ and $V_a$, respectively, and $k_A$ and $k_a$ are
Kac-Moody levels of $U(1)_A$ and $G_a$, respectively. And the square of
the gauge coupling is written in terms of the inverse of the vacuum 
expectation value (VEV) of the dilaton, i.e., $k_a\VEV{S}=1/g_a^2$.

The parameter $\delta_{GS}$ in Eq.~(\ref{trans}) is related to the
conditions for the anomaly cancellations,\footnote
{ 
$C_a\equiv\Tr_{G_a} T(R)\,Q_A$. $T(R)$ is the Dynkin index of the
representation $R$, and we use the convention that $T(\mbox{fundamental
rep.}) = 1/2$.
}
\begin{equation}
2\pi^2\delta_{GS}\,= \,\frac{C_a}{k_a}
\,= \,\frac{1}{3k_A}\tr {Q_A}^3 
\,= \,\frac{1}{24}\tr Q_A.
\end{equation}
The last equality is required by the cancellation of the mixed
gravitational anom-aly. These anomaly cancellations are understood in
the context of the Green-Schwarz mechanism \cite{GS}.  

If there is another $U(1)'$ gauge symmetry, the additional condition
\begin{eqnarray}
\Tr Q_A^2 Q' = 0
\label{mixed_anom}
\end{eqnarray}
is required, since the mixed anomaly $U(1)_A^2 U(1)'$ cannot be canceled
by the nonlinear transformation of the dilaton. Moreover, the coupling
unification $g_3=g_2=\sqrt{\frac{5}{3}}g_Y$ requires the relations
\begin{eqnarray}
k_3=k_2=\frac{3}{5}k_Y 
\;\;\;\Rightarrow\;\;\;
C_3=C_2=\frac{3}{5}C_Y. 
\label{GUT_relation}
\end{eqnarray}
The conditions (\ref{mixed_anom}) and (\ref{GUT_relation}) are
automatically satisfied in the case that the anomalous $U(1)$ charges
respect the $SU(5)$ GUT symmetry. 

One of the most interesting features of the anomalous $U(1)$ gauge
symmetry is that it induces the Fayet-Iliopoulos $D$-term (F-I term)
radiatively \cite{U(1)}. Since the K\"ahler potential $K$ for the
dilaton $S$ must be a function of $S+S^\dagger-\delta_{GS}V_A$ for the
$U(1)_A$ gauge invariance, the F-I term can be given as
\begin{equation}
\int d^4\theta \,K(S+S^\dagger-\delta_{GS}V_A)
=  \left(-\frac{\delta_{GS}K'}{2}\right)D_A + \cdots
  \equiv \xi^2 D_A +\cdots,
\label{FIterm}
\end{equation}
where we take the sign of $Q_A$ so that $\xi^2 >0$. 
  
When some superfields $\Phi_i$ have anomalous $U(1)$ charges $q_i$, the
scalar potential becomes\footnote
{
Throughout this paper we denote both superfields and their
scalar components with uppercase letters.
}
\begin{eqnarray}
V_{\mbox{\scriptsize scalar}}
&= &\frac{g_A^2}{2}
      \left(\sum_i q_i |\Phi_i|^2 
            +\xi^2 \right)^2,        
\end{eqnarray}
where $1/g_A^2=k_A \VEV{S}$. If one superfield has a negative anomalous
$U(1)$ charge, it takes the vacuum expectation value (VEV). Below we
assume the existence of the field $\Phi$ with a negative charge and
normalize the anomalous $U(1)$ charges so that $\Phi$ has a charge
$-1$. In this case the VEV of the scalar component $\Phi$ is given by
\begin{equation}
\VEV{\Phi}=  \xi \equiv \lambda \Mp, 
\label{VEV_Phi}
\end{equation}
which breaks the anomalous $U(1)$ gauge symmetry. ($\Mp$ is some gravity 
scale and usually taken as the reduced Planck mass, $1/\sqrt{8\pi
G_N}$.)

We first discuss fermion masses. In general, the Yukawa hierarchy can be
explained by introducing a flavor dependent $U(1)$ symmetry
\cite{FN,yukawa,DLLRS}. We can adopt the anomalous $U(1)$ gauge symmetry 
as the above $U(1)$ symmetry. Suppose that the standard model matter 
fields $Q_i$, $U^c_i$, $D^c_i$, $L_i$, $E^c_i$, $H_u$ and $H_d$ have the 
anomalous $U(1)$ charges $q_i$, $u_i$, $d_i$, $l_i$, $e_i$, $h_u$ and 
$h_d$, respectively, which are assumed to be non-negative integers. 
If the field $\Phi$ with charge $-1$ is a singlet under the standard 
model gauge symmetry, the superpotential can be written as
\begin{eqnarray}
W \sim
  \left(\frac{\Phi}{\Mp}\right)^{q_i+u_j+2h_u}
    H_uQ_i U^c_j + \cdots. 
\end{eqnarray}
Since the scalar component of $\Phi$ has the VEV in Eq.~(\ref{VEV_Phi}), 
we obtain the hierarchical mass matrices
\begin{eqnarray}
(M_u)_{ij}
&\sim&
\lambda^{q_i+u_j+h_u} \VEV{H_u}
= V^u_L 
\left(
\begin{array}{ccc}
m_u & & \\
 & m_c & \\
 & & m_t
\end{array}
\right)
V^{u\dagger}_R , 
\label{diag_matrices1}\\
(M_d)_{ij}
&\sim&
\lambda^{q_i+d_j+h_d} \VEV{H_d} 
= V^d_L 
\left(
\begin{array}{ccc}
m_d & & \\
 & m_s & \\
 & & m_b
\end{array}
\right)
V^{d\dagger}_R,
\label{diag_matrices2}
\end{eqnarray}
where $V^{u,d}_{L,R}$ are $3\times 3$ unitary diagonalizing matrices, and
$(V^u_L)_{ij} \sim \lambda^{|q_i-q_j|}$, $(V^u_R)_{ij} \sim
\lambda^{|u_i-u_j|}$ and so on. The diagonalized masses of quarks,
$m_f$, are given as follows:
\begin{eqnarray}
(m_u)_i \sim \lambda^{q_i+u_i+h_u} \VEV{H_u} 
\;\;\; {\rm and} \;\;\;
(m_d)_i \sim \lambda^{q_i+d_i+h_d} \VEV{H_d} .
\label{fermion_mass}
\end{eqnarray}
The Cabbibo-Kobayashi-Maskawa matrix is
\begin{eqnarray}
V_{\rm CKM}
= V^u_L {V^d_L}^{\dagger}
\sim \left(
\begin{array}{ccc}
1&\lambda^{|q_1-q_2|}&\lambda^{|q_1-q_3|}\\
\lambda^{|q_2-q_1|}&1&\lambda^{|q_2-q_3|}\\
\lambda^{|q_3-q_1|}&\lambda^{|q_3-q_2|}&1 
\end{array}
\right) ,
\end{eqnarray}
which is determined only by the charges of the left-handed quarks,
$q_i$. The relation $V_{12}V_{23}\sim V_{13}$ is naturally understood with
this mechanism, and if we take $q_i= (3,2,0)$ and $\lambda \sim 0.2$, we
can reproduce the measured value.

If there are right-handed neutrinos $N_i^c$ with $U(1)_A$ charges $n_i$,
Dirac and Majorana neutrino masses are given by
\begin{eqnarray}
  (M^D)_{ij}&\sim& \lambda^{l_i+n_j+h_u}\VEV{H_u},\\  
  (M^M)_{ij} &\sim& M_m \lambda^{n_i+n_j}.
\end{eqnarray}
Through the see-saw mechanism \cite{seesaw} the left-handed neutrino
mass matrix is given by\footnote
{
Here we have to introduce a Majorana mass scale $M_m$ that is smaller than
$\Mp$. If we simply take $M_m\sim\Mp$, the neutrino mass becomes
$O(10^{-5}{\rm eV})$, which is smaller than the values indicated by
various experiments.
}
\begin{eqnarray}
  (M_\nu)_{ij} &\sim& \lambda^{l_i+l_j+h_u}\frac{\VEV{H_u}^2}{M_m}.
\end{eqnarray}
The mixing matrix for the lepton sector \cite{MNS} is induced as for the
quark sector:
\begin{eqnarray}
V_{\rm MNS}
= V^\nu_L {V^e_L}^{\dagger}
\sim \left(
\begin{array}{ccc}
1&\lambda^{|l_1-l_2|}&\lambda^{|l_1-l_3|}\\
\lambda^{|l_2-l_1|}&1&\lambda^{|l_2-l_3|}\\
\lambda^{|l_3-l_1|}&\lambda^{|l_3-l_2|}&1 
\end{array}
\right).
\end{eqnarray}
This matrix is also determined only by the charges of the left-handed leptons,
$l_i$. If we take $l_i=(4,2,2)$, it gives mixing angles which are
indicated by the MSW small angle solution for the solar neutrino problem and
the large angle solution for the atmospheric neutrino anomaly
\cite{SuperK,neutrino,coset}. Other charge assignments may result in mixing
angles and masses which are consistent with the other solutions and/or the LSND
experiment, but we will not discuss it further.

If there is no massless superfield with a negative $U(1)_A$ charge, SUSY
is broken spontaneously. For example, suppose that all fields have
non-negative anomalous $U(1)$ charges except the field $\Phi$ and that
$\Phi$ has the superpotential\footnote
{
Note that $m$ has a charge $-2$. Here we only assume that $m$ is
generated by some unknown strong dynamics. Later we will show this
concretely.
}
\begin{equation}
W= \frac{1}{2}m\Phi^2.
\end{equation}
From the scalar potential 
\begin{eqnarray}
V
_{\mbox{\scriptsize scalar}} 
= \frac{1}{2}g_A^2 
   \left( \sum_i q_i |\Phi_i|^2 - |\Phi|^2  + \xi^2 \right)^2   
   + m^2 |\Phi|^2, 
\end{eqnarray}
it can be seen that at the global minimum point,
$\VEV{\Phi} =  \sqrt{\xi^2-{m^2}/{g_A^2}}$
and $\VEV{\Phi_i} = 0$,
\begin{eqnarray}
\VEV{F_\Phi}= -m\sqrt{\xi^2-\frac{m^2}{g_A^2}}\;,\;\; 
\VEV{D_A}= -m^2.
\end{eqnarray}
In this case SUSY is broken spontaneously. The soft scalar mass squared $m^2_i$
for the field $\Phi_i$ is induced through the $D$-term, and it is proportional
to the anomalous $U(1)$ charge $q_i$: 
\begin{equation}
m^2_i = -q_i \VEV{D_A}.
\end{equation} 

The flavor-dependent $U(1)_A$ charges, which are needed for solving the
fermion mass hierarchy, inevitably induce non-degenerate scalar fermion
masses, which cause a large contribution to FCNC processes. Therefore
we usually adopt a decoupling scenario in which the soft SUSY breaking
scalar masses for the first two generations are much larger than the
weak scale, in order to suppress the FCNC process, while the masses for
the third generation and the gauginos are as large as the weak scale for
"naturalness" \cite{DG,CKN}. However, it has been pointed out by Arkani-Hamed
and Murayama that large soft scalar masses for the first two generations
tend to drive the stop mass squared to a negative value at the two loop level,
so that this scenario is problematic 
\cite{neg_stop}.\footnote
{
One way to suppress these unwanted two loop contributions is to
introduce extra vector-like quarks at heavy squark mass scales
\cite{HKN}.
}
In the next section we discuss another scenario that avoids this problem
without discarding the idea that the anomalous $U(1)$ gauge symmetry
accounts for both SUSY breaking and the fermion mass hierarchy simultaneously.

\section{$F$-term contribution of the dilaton}
Recently, Arkani-Hamed, Dine and Martin \cite{ADM} pointed out that the
$F$-term  contribution of the dilaton to the SUSY breaking parameters
cannot be neglected, especially when the dilaton is stabilized by the
deformation of the K\"ahler potential. This implies that the phenomenology
in anomalous $U(1)$ SUSY breaking models can change significantly. This is
because the $F$-term of the dilaton gives gaugino masses, through which
flavor-independent sfermion masses are generated by loop
corrections. Since flavor-dependent sfermion masses come from the
$D$-term contribution, the magnitude of the flavor violation is
controlled by the ratio of the $D$-term to the $F$-term. In this
section we show that the order of the ratio is determined under some
assumptions.\footnote{
In Refs.~\cite{BCCM,Irges}, phenomenological aspects 
under different assumptions from ours are discussed.
}

First we review their argument using an explicit example. As we have seen,
when some strong dynamics induces the effective mass term of the field
$\Phi$ with the $U(1)_A$ charge $-1$, SUSY is broken dynamically. As the
strong dynamics Arkani-Hamed, Dine and Martin adopted the $SU(N_c)$ gauge 
theory with one flavor
$Q$ and $\bar Q$, which have anomalous $U(1)$ charges $q$ and $\bar q$,
respectively. The superpotential in tree level is
\begin{equation}
W= C  \left(\frac{Q\bar Q}{\Mp^2}\right)
  \left(\frac{\Phi}{\Mp}\right)^{q+\bar q}\Mp^3.
\end{equation}
Below the dynamical scale $\Lambda$ of the $SU(N_c)$ gauge theory, the
effective superpotential with the canonically normalized meson
superfield $t \equiv \sqrt{2Q\bar Q}$ becomes
\begin{equation}
W= C \left(\frac{t^2}{2\Mp^2}\right)
  \left(\frac{\Phi}{\Mp}\right)^{q+\bar q}\Mp^3
   +(N_c-1)\left(\frac{2\Lambda^{3N_c-1}}{t^2}\right)^{\frac{1}{N_c-1}}.
\end{equation}
The second term is the Affleck, Dine and Seiberg (ADS) superpotential
$W_{\rm np}$ \cite{ADS}, through which the superpotential $W$ depends on 
the dilaton $S$. That is,\footnote
{
$\left({\Lambda}/{\Mp}\right)^{3N_c-1}= e^{-8\pi^2k_N S}$, 
where $k_N$ is the Kac-Moody level of the $SU(N_c)$ gauge group.
}
\begin{equation}
W_{\rm np} \propto e^{-({8\pi^2 k_N S})/({N_c-1})} 
 = e^{-S/\delta},
\end{equation}
where $\delta=(N_c-1)/(8\pi^2k_N) \ll 1$. This small parameter $\delta$
plays an important role in the following. The total K\"ahler
potential is $t^\dagger t(e^{2qV^A}+e^{2\bar q V^A})/2 + \Phi^\dagger
\Phi e^{-2V^A} + K(S+S^\dagger - \delta_{GS} V^A)$. Then the scalar
potential is 
\begin{equation}
V= K''|F_S|^2
     +\left|\frac{\partial W}{\partial\Phi}\right|^2
     +\left|\frac{\partial W}{\partial t}\right|^2
     +\frac{1}{2g_A^2}D_A^2,
\end{equation}
where $F_S= -{W^\dagger}'/K^{\prime \prime}$ and $D_A=
-g_A^2\left(\frac{q+\bar q}{2}|t|^2-|\Phi |^2+\xi^2\right)$.

If the K\"ahler potential for the dilaton $K$ is given by
\begin{equation}
K_{tree}= -\ln (S+S^\dagger ),
\end{equation}
which can be induced by a stringy calculation at tree level, the potential
for the dilaton has a run-away vacuum. A solution for dilaton
stabilization is to deform the K\"ahler potential. For example, we take,
following Ref.~\cite{ADM},
\begin{equation}
K= -\ln (S+S^\dagger)-\frac{2s_0}{S+S^\dagger}
   +\frac{b+4s_0^2}{6(S+S^\dagger)^2},
\label{Kahler_ADM}
\end{equation}
where $s_0$ and $b$ are non-negative constants. If $b < \delta^2$, there
is a stable local minimum at $\bar s \equiv s_0-\delta$. At that
minimum, the ratio of the VEV of $D_A$ and $F_S$ can be calculated as
\begin{equation}
r\equiv \frac{|\VEV{D_A}|}{|\VEV{F_S}|^2}= \frac{3\delta}{2s_0^3}.
\end{equation}
Note that the ratio $r$ can be small; i.e., the $F$-term of the dilaton
can dominate the ordinary $D$-term contribution. Therefore the
phenomenology is significantly changed. Assuming a canonical gauge kinetic
function, the gaugino masses are
\begin{equation}
M_{{1}/{2}}= \frac{\VEV{F_S}}{\VEV{S+S^\dagger}},
\end{equation}
and the gravitino mass is
\begin{eqnarray}
m_{3/2}= {\frac{1}{\sqrt{3}}}\sqrt{K''}\VEV{F_S}.
\end{eqnarray}
On the other hand, the scalar fields with the anomalous $U(1)$ charges
$q_i$ obtain the soft masses
\begin{equation}
m_i^2= -q_i\VEV{D_A}
      +\frac{1}{3}K'' |\VEV{F_S}|^2
      +\cdots.
\end{equation}
Since $K''(\bar s) \sim \delta^2$ and $r \sim \delta$, we have $M_{1/2}^2 :
m^2_i : m^2_{3/2} \sim 1 : \delta : \delta^2$; that is, $M_{1/2}^2 \gg
m^2_i \gg m^2_{3/2}$. These are  so-called no-scale-like SUSY breaking
parameters. This relation is given at the Planck scale. At the weak
scale the contribution from the loop corrections of the gauginos can be
the main part of the scalar masses.

In the above argument we have assumed a specific K\"ahler potential
and considered only the case with one flavor. What we emphasize in this
paper is that the ratio $r$ is generally determined to be of the
order of $0.01$ under some assumptions discussed below.

To begin with, we extend the number of flavors from $1$ to $N_f$
($<\!\!N_c$), $Q^i$, $\bar Q_j$ ($i,j= 1\cdots N_f$) and increase the
number of pairs $Q\bar Q$ that couple to the field $\Phi$ in the
superpotential. The fields $Q^i$ and $\bar Q_j$ have anomalous $U(1)$
charges $q_i$ and $\bar q_j$, respectively. The superpotential in tree
level is
\begin{equation}
W= C^{\bar \imath_1, \cdots , \bar \imath_n}_{i_1, \cdots , i_n}
  \left(\frac{Q^{i_1}\bar Q_{\bar \imath_1}}{\Mp^2}\right)\cdots
  \left(\frac{Q^{i_n}\bar Q_{\bar \imath_n}}{\Mp^2}\right)
  \left(\frac{\Phi}{\Mp}\right)^{q_{sum}}\Mp^3,
\label{general_W}
\end{equation}
where
\begin{equation}
q_{sum}= \sum_{p= 1}^n \left( q_{i_p}+{\bar q}_{\bar \imath_p}\right).
\end{equation}

In this generalized case the ratio $r= |D_A|/|F_S|^2$ is also calculated
explicitly (see the Appendix):
\begin{equation}
r= \frac{8\pi^2k_N}{N_c-N_f+\frac{N_f}{n}}
  \left(\frac{K''}{-K'}\right)
 +\left\{\frac{2C_N}{N_c-N_f+\frac{N_f}{n}}-1\right\}
  \left(\frac{K''}{-K'}\right)^2.
\label{general_r}
\end{equation}
Note that the order of the ratio $r$ is almost determined by the ratio
${K''}/{K'}$. In the following subsection we discuss the assumptions under
which the ratio ${K''}/{K'}$ (that is, $r$) is generally determined.

Our first assumption is that the dilaton is stabilized by corrections to
the tree level K\"ahler potential; that is, K\"ahler potential $K$ is
made up of the tree level K\"ahler potential $K_{tree}=
-\ln(S+S^\dagger)$ and  some correction $K_{cor}$, which goes to zero
in the  weak coupling limit.
\begin{equation}
K= K_{tree}+K_{cor}.
\end{equation}
Since there is a non-perturbative superpotential $W_{\rm np}$,
\begin{equation}
W_{\rm np} \propto e^{-S/\delta},
\label{Wnp_delta}
\end{equation}
where $\delta= (N_c-N_f)/(8\pi^2 k_N) \ll 1$, the condition for the
stabilization of the dilaton is
\begin{equation}
\frac{K'''}{K''}
= \frac{K'''_{tree}+K'''_{cor}}{K''_{tree}+K''_{cor}}
= -\frac{1}{\delta}.
\label{Cond_stabilization}
\end{equation}
Here we have considered only the non-perturbative superpotential $W_{\rm
np}$. Indeed there is the dilaton-dependent part of the tree level
superpotential, but its presence does not change the following argument
significantly.

In the analysis in this paper, we take $\VEV{S}\sim 2$ to realize the
standard gauge coupling at the GUT scale $g_a^{-2}=\VEV{S}\sim 2$. Since
\begin{eqnarray}
\frac{K'''_{tree}}{K''_{tree}}=\frac{S+S^\dagger}{3}= O(1), 
\end{eqnarray}
we can consider two cases which satisfy the condition
(\ref{Cond_stabilization}).
\begin{description}
\item[{\rm case 1,}] large numerator :
\begin{equation}
K'''_{cor} \gg K'''_{tree}
\;\;\;{\rm and}\;\;\; 
K'''_{cor} \gg K''_{cor}, K''_{tree}
\end{equation}
\item[{\rm case 2,}] small denominator :
\begin{equation}
K'''_{cor} \le K'''_{tree} 
\;\;\;{\rm and}\;\;\;
K''_{tree}+K''_{cor} \ll K'''_{tree} 
\end{equation}
\end{description}
Case 1 corresponds to the K\"ahler potential discussed in
Ref.~\cite{BCCM,Irges}\footnote
{
In Ref.~\cite{BCCM} it is insisted that case 2 is inconsistent with a
zero cosmological constant, but the authors of that paper 
ignore a constant term of the
superpotential. Including this constant term, case 2 is also consistent
\cite{KNOW}.
}. 
Here we consider case 2 only. (This is the second assumption.) In this
case we are to expect that the cancellation between $K''_{tree}$ and
$K''_{cor}$ occurs,
\begin{figure}
\centerline{\epsfxsize=12cm \epsfbox{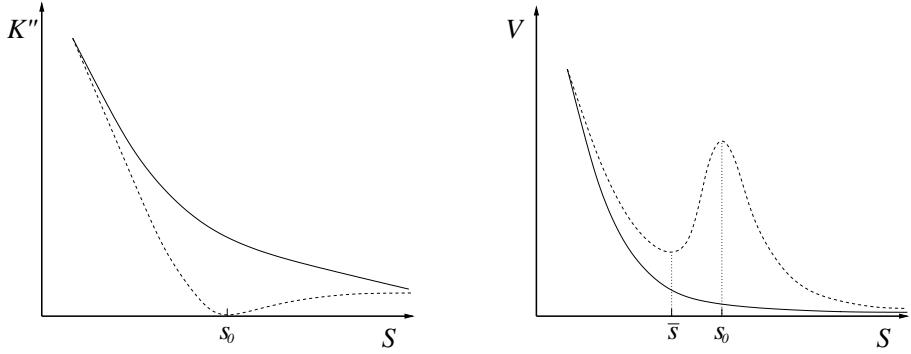}}
\caption{These figures show how the deformation of the K\"ahler
potential stabilizes the potential of the dilaton. The solid lines are 
for the tree level K\"ahler potential, and the dashed lines are for
the deformed K\"ahler potential}
\label{fig_kahler}
\end{figure}
that is, that the K\"ahler potential behaves like Fig.~\ref{fig_kahler}
and has a local minimum at $s_0$. Therefore $K''$ can be expanded around the
local minimum $s_0$ as
\begin{equation}
 K''(S) =  \alpha + \frac{1}{2}K''''(s_0)(S+S^\dagger-2s_0)^2+\cdots.
\end{equation} 
Then the condition of the existence of a local minimum is $\alpha <
O(\delta^2)$, and at $\bar s= s_0-\delta$ the potential is minimized. We
find the ratio at the local minimum $\VEV{S}=\bar s \sim 2$,
\begin{equation}
\frac{K''(\bar s)}{K'(\bar s)}= 2\delta^2\frac{K''''(s_0)}{K'(s_0)}.
\end{equation}
We have no exact information concerning $K''''(s_0)$ and $K'(s_0)$.
Therefore we
estimate the ratio $K''''(s_0)/K'(s_0)$ by the ratio at tree level,
$K''''_{tree}(s_0)/K'_{tree}(s_0)$. (This is the third assumption.) This
is because we expect that the cancellation between $K''_{tree}(s_0)$ and
$K''_{cor}(s_0)$ occurs only for the second derivative of the K\"ahler
potential, not for other derivatives. Therefore we rewrite
$K''''(s_0)/K'(s_0)$ as
\begin{equation}
\frac{K''''(s_0)}{K'(s_0)}=\frac{K''''_{tree}(s_0)}{K'_{tree}(s_0)}
    \times \gamma = \frac{3}{4 \bar s^3} \times \gamma.
\end{equation} 
In general $\gamma$ is dependent on an explicit form of the K\"ahler
potential, but we assume that this dependence is order unity and less
important. For example, $\gamma =1$ when we take the form of
Eq.~(\ref{Kahler_ADM}) as the K\"ahler potential. We finally find
\begin{equation}
\frac{K''(\bar s)}{K'(\bar s)}
= -\frac{3\delta^2}{2\bar s^3}\times\gamma.
\;\;\;( \gamma \sim O(1))
\end{equation}
As we have remarked, the condition (\ref{Cond_stabilization}) is an
approximate one. Following the exact condition (\ref{exact_ratio})
derived in the Appendix, $\delta$ must be changed slightly as follows:
\begin{equation}
\delta =  \frac{N_c-N_f+\frac{N_f}{n}}{8\pi^2 k_N}.
\label{exact_delta}
\end{equation}
Finally, using $\delta$ in Eq.~(\ref{exact_delta}) the ratio $r$ is given
by
\begin{equation}
r= \frac{3\delta}{2\bar s^3} \times \gamma.
\end{equation}

It should be noted that in this scenario the order of $r$ is almost
determined by $\delta$, that is, the inverse of $8\pi^2$, so that $r\sim
0.01$. Since the ratio $r$ represents the magnitude of the flavor
violation, the model-independent nature of $r$ implies that we can make
model-independent predictions for flavor violating processes. In the
next section, we discuss FCNC processes using the ratio $r$. 

\section{Lepton flavor violation}
The flavor-independent part of the sfermion masses at the weak scale is
induced by loop corrections of gauginos, which are easily calculated 
using the renormalization group equations (RGE). Generally, the loop
corrections to the slepton masses are smaller than those to the squark
masses. Therefore the flavor-violating effects of the non-degeneracy
among the sfermion masses are larger in lepton flavor violating
processes like $\mu\to e\gamma$ than baryon flavor violating processes
like $K^0\bar K^0$ mixing. In this paper we only discuss flavor
violation from the non-degeneracy between the first and the second
generations, because experimental constraints on flavor violation
 are the most stringent. 
Therefore we neglect the third generation below.\footnote
{
In the case that the second and the third generations have the same
$U(1)_A$ charges, we must take into account the flavor violation from
the non-degeneracy of the first and the third generations. However, this
violation can be at most of the same order as that between the first and the
second generations, and its contribution is not always additive. 
} 

We first calculate the sfermion masses induced by RGE. Assuming that the
SUSY breaking parameters are given around the GUT scale and that below
the GUT scale the model is the minimal SUSY standard model, the sfermion
masses $m_{\tilde f_i }^2$ around the SUSY breaking scale can be written
as follows in terms of the sfermion mass squares $m^2_{\tilde f_i 0}=-f_i
\VEV{D_A}$ and the universal gaugino mass $M_{1/2}$ around the GUT scale:
\begin{eqnarray}
m^2_{\tilde f_i} &= & m^2_{\tilde f_i 0} + \xi_{\tilde f} M_{1/2}^2,
\label{sfermion_mass}
\end{eqnarray}
where the coefficients $\xi_{\tilde f}$ are estimated numerically
as tabulated below.
\begin{eqnarray}
\begin{array}{|c|ccccc|}
\hline
\tilde f & \tilde q_L & \tilde u_R & \tilde d_R & \tilde l_L &
\tilde e_R \\
\hline
\xi_{\tilde f} & 6.5 & 6.1 & 6.0 & 0.52 & 0.15 \\
\hline
\end{array}
\nonumber 
\end{eqnarray}

The sfermion masses in Eq.~(\ref{sfermion_mass}) are given in the base
of the anomalous $U(1)$ gauge symmetry. On the other hand, it is
convenient to discuss the flavor violating processes in terms of
off-diagonal elements of the sfermion mass matrices in the basis in which
fermion mass matrices are diagonal. Because the diagonalizing matrices
are given by 
\begin{eqnarray}
 V_f \sim 
\left(
\begin{array}{cc}
1 & \lambda^{\Delta f}\\
-\lambda^{\Delta f} & 1
\end{array}
\right)
\end{eqnarray}
in Eqs.~(\ref{diag_matrices1}) and (\ref{diag_matrices2}), 
off-diagonal elements of the sfermion masses are
\begin{eqnarray}
\delta m^2_{\tilde f} \sim \Delta f\lambda^{\Delta f} \VEV{D_A},
\end{eqnarray} 
where $\Delta f = |f_1-f_2|$. Note that no flavor violation occurs in
the case $\Delta f=0$, because the sfermion masses are degenerate. 
Maximal flavor violation occurs in the case that $\Delta f=1$ for
$\lambda \sim 0.2$.

Using the normalized off-diagonal elements of sfermion masses,
$\delta_{\tilde f} \equiv \delta m^2_{\tilde f}/m^2_{\tilde f}$, the
constraints from FCNC processes can be roughly written as
\begin{equation}
\sqrt{\delta_{\tilde q} \delta_{\tilde d}}
 < \frac {m_{\tilde q}}{500 {\rm GeV}}
\times 10^{-3}
\label{constraint_KK1}
\end{equation}
from the $K^0\bar K^0$ mixing, and
\begin{equation}
\delta_{\tilde l},\;\delta_{\tilde e}
< \left(\frac{m_{\tilde l,\;\tilde e}}
{100 {\rm GeV}}\right)^2
\times 10^{-3}
\label{constraint_LFV1}
\end{equation}
from the $\mu \to e \gamma$ process for moderate $\tan\beta$
\cite{FCNC}. 

Here we define the ratio $R$ between the gaugino mass squared and the
non-degenerate mass squared around the GUT scale instead of $r$:
\begin{eqnarray}
R\equiv\frac{\VEV{D_A}}{M^2_{1/2}}= 4\bar s^2 r
=  \frac{6\delta}{\bar s}\times \gamma.
\end{eqnarray}
Because $r$ is generally determined as about $0.01$, as stated in the 
previous section, and $\bar s = 1/g^2 \sim  2$, we have $R\sim 0.1$. 
Using $R$ and $\xi_{\tilde f}$,
 the normalized off-diagonal elements are written
\begin{eqnarray}
\delta_{\tilde f} \sim \Delta f \lambda^{\Delta f} 
                    \frac{R}{\xi_{\tilde f}}.
\end{eqnarray}
Then the constraints (\ref{constraint_KK1}) and (\ref{constraint_LFV1})
become
\begin{eqnarray}
\frac{R}{\xi_{\tilde q}^{3/2}} 
(\Delta q \Delta d)^{1/2} \lambda^{(\Delta q+\Delta d)/2} 
&<& \frac{M_{1/2}}{500{\rm GeV}}\times 10^{-3},
\label{constraint_KK2}
\\
\frac{R}{\xi_{\tilde l}^2} \Delta l \lambda^{\Delta l},
\;\;
\frac{R}{\xi_{\tilde e}^2} \Delta e \lambda^{\Delta e}
&<& \left(\frac{M_{1/2}}{100{\rm GeV}}\right)^2
\times 10^{-3}.
\label{constraint_LFV2}
\end{eqnarray}

For example, if we assume $U(1)_A$ charges such that $\Delta f
=1$, which gives the most stringent constraints, the constraint from the
$K^0\bar K^0$ mixing requires that $M_{1/2} > 500$GeV, while the
constraint from the $\mu \to e \gamma$ process requires that $M_{1/2} >
3$TeV. This rough estimation of the $\mu \to e\gamma$ process is
too severe. Actually the constraint becomes weak when $M_{1/2}>1$TeV,
because we neglect the contribution to the slepton masses from the
$D$-term.

As the other extreme case, we could consider $\Delta f=0$, which results
in no flavor violation. 
However, the $U(1)_A$ charges, $f_i$, are related to the
fermion masses according to Eq.~(\ref{fermion_mass}). For instance, $m_e
/m_{\mu}\sim \lambda^{\Delta l +\Delta e}$, that is, $\Delta l +\Delta e
\sim 3$. This implies that we cannot take $\Delta l$ and $\Delta e$ to be 
zero simultaneously. Therefore the weakest constraint from the $\mu \to
e\gamma$ process is given in the case that ($\Delta l$, $\Delta e$) $=$
($3$, $0$) or ($0$, $3$), where the constraint becomes $M_{1/2}
>300$GeV. This is because $\Delta l \lambda^{\Delta l} = 0.2 \to 3
\times 0.2^3$, so that the amplitude decreases by one order. In the next
section we report the results of numerical calculations using 
the $U(1)_A$ charge
assignments corresponding to the above two extreme cases.

Before considering these calculations, 
note that it has been assumed implicitly
in the above argument that there are no other contributions to the
sfermion masses. Is it true? Are there any other contributions?
If there is another flavor-independent contribution, the constraint
becomes weak. Indeed two other flavor-independent contributions to
sfermion masses can be considered.

One contribution is induced by GUT interactions. Assuming the GUT,
slepton masses receive much larger loop corrections from the GUT
interactions than from only the weak interactions. The other 
correction comes from a K\"ahler potential,
\begin{eqnarray}
  K= Z_{i\bar\jmath}(S,S^\dagger)\Phi_i \Phi_{\bar \jmath}^\dagger.
\end{eqnarray}
which contributes to the soft scalar masses as
\begin{eqnarray}
 \Delta m^2_{i\bar\jmath}
  &=&  |F_S|^2 \frac{\del^2 Z_{i\bar\jmath}}
           {\del S \del S^\dagger}.
\end{eqnarray}
Though we do not know a definite reason that the dilaton has 
flavor-independent interactions with chiral superfields, we calculate the
branching ratio in the next section under the assumption that $\Delta
m^2_{i\bar\jmath} = M^2_{1/2} \delta_{i \bar \jmath}$.\footnote{
In Ref.~\cite{GLLV}, lepton flavor violating processes are calculated
in the presence of general non-diagonal contributions.
}
Moreover, we
should consider the $F$-term contributions of the other moduli.
However, it
is not so unnatural to assume $F_S \gg F_M$, because a large $F_S$ is
realized in the special case that the second derivative of the K\"ahler
potential is extremely small. In the next section we consider two
cases with these contributions. 

Before ending this section, we should comment on CP violation. The
experimental value of the CP violation in $K^0\bar K^0$ mixing results
in one order severer constraint on the gaugino mass than that resulting
from the real part of the mixing. 
However, since we do not know the origin of the CP violation, 
we do not have to consider it seriously in deriving
conservative constraints.

\section{Numerical calculations}
In this section we report on numerical calculations for the $\mu \to e\gamma$
process\cite{LFV,cal_LFV,cancel}. If we know all parameter values at the GUT
scale, we can get the values at the low energy scale. Then we can
estimate the branching ratio of $\mu \to e \gamma$ by using the exact
calculation in Ref.~\cite{cal_LFV}. There are five kinds of 
free parameters in MSSM at the GUT scale, the sfermion masses
$m^2_{\tilde f_i 0}$, the universal gaugino mass $M_{1/2}$, the
trilinear parameters $A$, the higgsino mass $\mu$ and the Higgs mixing
parameter $B$.

The sfermion masses are given by the $U(1)_A$ charges, the ratio $R$ and
gaugino mass $M_{1/2}$. In the following calculations, $R$ and $M_{1/2}$
are taken as free parameters. As the assignments of the anomalous $U(1)$
charges, we consider two types in Table \ref{charge_assign}. Type A
and Type B correspond to the case that gives the most stringent
constraint and the case that gives the weakest constraint, respectively, 
as discussed in the previous section. Type A respects the $SU(5)$ GUT
symmetry and reproduces the quark and lepton mass matrices well. Type B
was first discussed in Ref.~\cite{NW}, but it does not respect the $SU(5)$ GUT
symmetry, so that it is not trivial to satisfy the conditions
(\ref{mixed_anom}) and (\ref{GUT_relation}), and it does not explain the
largeness of the neutrino mixing angle.
\begin{table}[h]
\begin{eqnarray}
\begin{array}{|c|cccccc|}
\hline
 & q_i & u_i & e_i & d_i & l_i & h_u, h_d\\
\hline
{\rm Type~A} & 3,2,0 & 3,2,0 & 3,2,0 & 4,2,2 & 4,2,2& 0,0\\
{\rm Type~B} & 3,2,0 & 4,1,0 & 4,1,0 & 3,3,2 & 3,3,2& 0,0\\
\hline
\end{array}
\nonumber
\end{eqnarray}
\caption{Assignments of the anomalous $U(1)$ charges.}
\label{charge_assign}
\end{table}
\\
Since trilinear parameters $A$ at the GUT scale are much smaller than
the gaugino mass $M_{1/2}$, they can be neglected and set to zero for
simplicity. The absolute value of the higgsino mass $\mu$ is determined
by the weak scale, and the parameter $B$ is determined by the ratio of
the VEV of the Higgs $\tan\beta$. (We use the tree potential in the
calculation.)

Therefore, as the free parameters in our scenario, we have the gaugino
mass $M_{1/2}$, the ratio $R=D_A/M_{1/2}$, the sign of $\mu$, and
$\tan\beta$. Below we calculate the branching ratios changing the values
of the first three parameters. We have considered only the case that
$R\sim 0.1$ in the previous section. However, since there are ambiguities
in $\gamma$ and some coefficients, we take the range of $R$ as $0.01
\sim 1$. We fix $\tan\beta=2$ in all calculations, because the
dependence of the $\mu\to e\gamma$ process on $\tan\beta$ is
simple. Since the dominant contribution is proportional to the
left-right mixing of slepton masses, the amplitude is proportional to
$\tan\beta$. For instance, if we take $\tan\beta=10$, the bound on the
gaugino mass $M_{1/2}$ is severer by a factor of $\sqrt{5}$.

Before giving the numerical results, we comment on the constraint
from the $K^0 \bar K^0$ mixing. The calculation in Ref.~\cite{FCNC} does
not contain the QCD corrections. In Ref.~\cite{QCD_cor1} it is shown
that the leading order QCD corrections tighten the constraints. However,
in their calculations, the hadronic scale $\mu_{had}$ is taken so that
$\alpha_s(\mu_{had})=1$. At such a scale perturbative calculations are
not reliable. Alternatively, we take $\mu_{had}=2$GeV, following to
Ref.~\cite{QCD_cor2}. In that paper Ciuchini et al. perform
the calculations containing the next-to-leading order and $B$-parameters 
given by lattice calculations. But their results are not much different
from the calculations with the vacuum-insertion-approximation (VIA) with
the leading-order QCD corrections (LO). Therefore we estimate the
constraints from the $K^0 \bar K^0$ mixing using VIA with the LO QCD
corrections.

\begin{figure}
\centerline{\epsfxsize=14cm \epsfbox{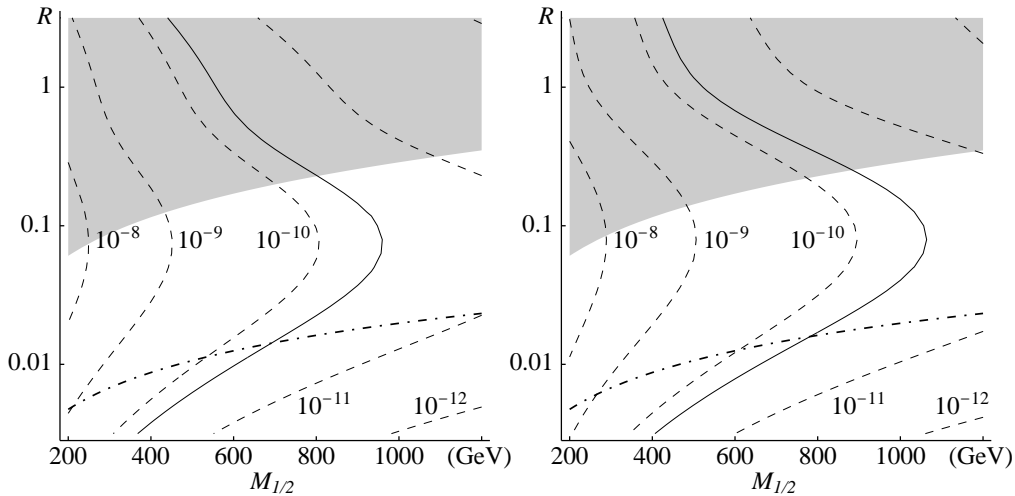}}
\caption{The branching ratio for $\mu\to e\gamma$ in
the Type A model with $\tan\beta=2$.
The left figure is for $\mu>0$ and the right for $\mu<0$.
The solid lines indicate the current experimental
bound ($4.9\times10^{-11}$). The left sides of the solid lines are
excluded. The shaded regions are excluded by the real part of the
$K^0 \bar K^0$ mixing. If the imaginary part of the contribution to the
$K^0 \bar K^0$ mixing is not suppressed, the excluded regions are
extended to the regions above the dot-dashed lines.}
\label{fig_br_A}
\end{figure}

Figure~\ref{fig_br_A} displays the branching ratio of $\mu \to e\gamma$
in the case that we use the charge assignment of Type A in Table
\ref{charge_assign} and only the loop corrections from the gauginos. Why
is there a peak around $R\sim 0.1$? The reason is as follows. Remember
that the branching ratio is proportional to $\delta m^2_{\tilde
e}/m^4_{\tilde e}\propto R/(\xi_{\tilde e} + e_R R)^2$. If the slepton
mass squared from the gaugino mass, $\xi_{\tilde e} M^2_{1/2}$,
dominates that from the $D$-term, $e_R R M^2_{1/2}$, we get a smaller
branching ratio for a smaller $R$. On the other hand, if the slepton
mass from the $D$-term dominates that from the gaugino mass, we get a
smaller branching ratio for a larger $R$. The border between two cases
must be the around $R\sim \xi_{\tilde e}/e_R \sim 0.08$. Here we use an
anomalous $U(1)$ charge for right-handed scalar muon $e_2=2$. This is
because the right-handed slepton gives the largest contribution to the
branching ratio. This is the reason for the fact that the branching
ratio takes its largest value around $R\sim 0.1$.

\begin{figure}
\centerline{\epsfxsize=14cm \epsfbox{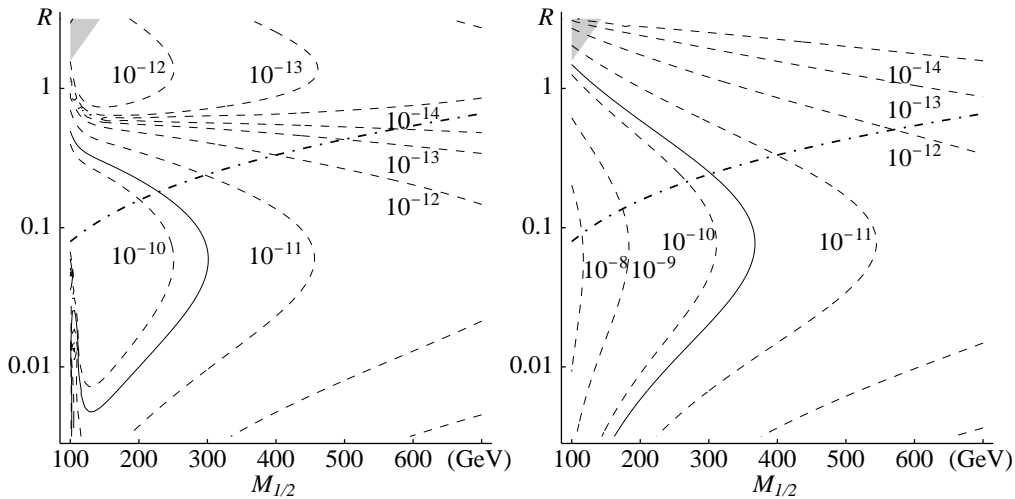}}
\caption{The branching ratio for $\mu\to e\gamma$ 
in the Type B model.
The meanings of the lines are the same as in Fig.~\ref{fig_br_A}.}
\label{fig_br_B}
\end{figure}

Figure~\ref{fig_br_B} shows the branching ratio in the case that we take
the charge assignment of Type B. Since the main contribution to the $K^0
\bar K^0$ mixing is proportional to $\sqrt{\delta_{\tilde q}
\delta_{\tilde d}}$, the constraint on the gaugino mass from the $K^0
\bar K^0$ mixing decreases by one order in the case that $\delta_{\tilde
d}=0$.\footnote
{
This is because the subdominant contribution to the $K^0\bar K^0$ mixing
gives the following weaker constraints on $\delta_{\tilde q}$ and
$\delta_{\tilde d}$ \cite{FCNC}:
\begin{eqnarray}
\delta_{\tilde q},\;\delta_{\tilde d}
< \frac{m_{\tilde q,\tilde d}}{500 {\rm GeV}} \times 10^{-2}.
\nonumber
\end{eqnarray}
} It can be seen that even the CP violation in $K$ meson does not give a
severe constraint. The figure for $\mu >0$ shows that there seems to be
a cancellation around $R\sim 1$. Since this charge assignment makes the
contribution from the left-handed sleptons vanish, the cancellation
between the Feynman diagrams of the right-handed scalar lepton, pointed
out in Ref.~\cite{cancel}, is realized. For $\mu <0$, the cancellation
occurs around $R\sim 3$, though it is not shown in the figure. Even for
this weakest constraint, near future experiments can prove the validity
of this scenario.

\begin{figure}
\centerline{\epsfxsize=14cm \epsfbox{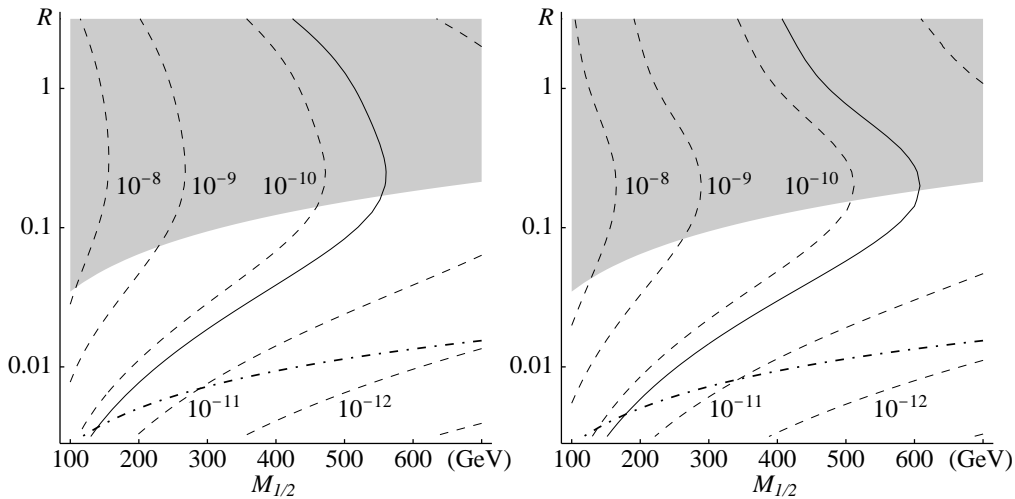}} 
\caption{The branching
ratio for $\mu\to e\gamma$ with the contribution from the
renormalization group from the Planck scale to the GUT scale. Here we
adopt the parameters of the minimal $SU(5)$ GUT. The anomalous $U(1)$
charges and the meanings of the lines are the same as in
Fig.~\ref{fig_br_A}.}  \label{fig_br_GUT}
\end{figure}

As discussed at the end of the previous section, if GUT exists, the
renormalization group from the Planck scale, $\Mp$, to the GUT scale,
$M_G$, contributes to the sfermion masses. In the case of $SU(5)$ GUT,
the contribution is
\begin{eqnarray}
m_{\tilde f}^2(M_G) 
  &= & m_{\tilde f}^2(\Mp)
      +\frac{C^f_5}{2b_5}M_{1/2}^2(M_G)
     \left[\left(\frac{\alpha_5(\Mp)}{\alpha_5(M_G)}\right)^2-1\right],
    \label{beta_depend}\\
  &\sim & m_{\tilde f}^2(\Mp)
         +\frac{\alpha_5(M_G)}{2\pi}C^f_5 M_{1/2}^2(M_G) 
          \ln \frac{\Mp}{M_G},
    \label{beta_independ}\\
  &\sim & m_{\tilde f}^2(\Mp)
         +0.03 \times C^f_5 M_{1/2}^2(M_G),
\end{eqnarray}
where $C^f_5$ is the coefficient of the quadratic Casimir ($C^f_5=72/5$
for the representation ${\bf 10}$ and $C^f_5=48/5$ for ${\bf 5^*}$), and
$b_5$ is the coefficient of the beta function
($b_5=3C(G)+T(R)$). Figure~\ref{fig_br_GUT} displays the branching ratio in
the minimal $SU(5)$ GUT case ($b_5= -3$). We can see that a smaller
gaugino mass, for example, $M_{1/2}=500$GeV, is allowed. Note that
though the amount of the contribution seems to depend on the coefficient
of the beta function in Eq.~(\ref{beta_depend}), there is no difference
between the minimal case and the non-minimal case ($b_5 > -3$) in the
approximation to first order in $\displaystyle \frac{\alpha_5}{2\pi}
\ln(\Mp/M_G)$, as in Eq.~(\ref{beta_independ}). 
Calculating $m_{\tilde f}^2(M_G)$ with no approximation,
it can be shown that $m_{\tilde f}^2(M_G)$ in the non-minimal cases
is slightly larger than that in the minimal case,
therefore the constraints on the gaugino mass in Fig.~\ref{fig_br_GUT} 
are weakened in the non-minimal cases.

\begin{figure}
\centerline{\epsfxsize=14cm \epsfbox{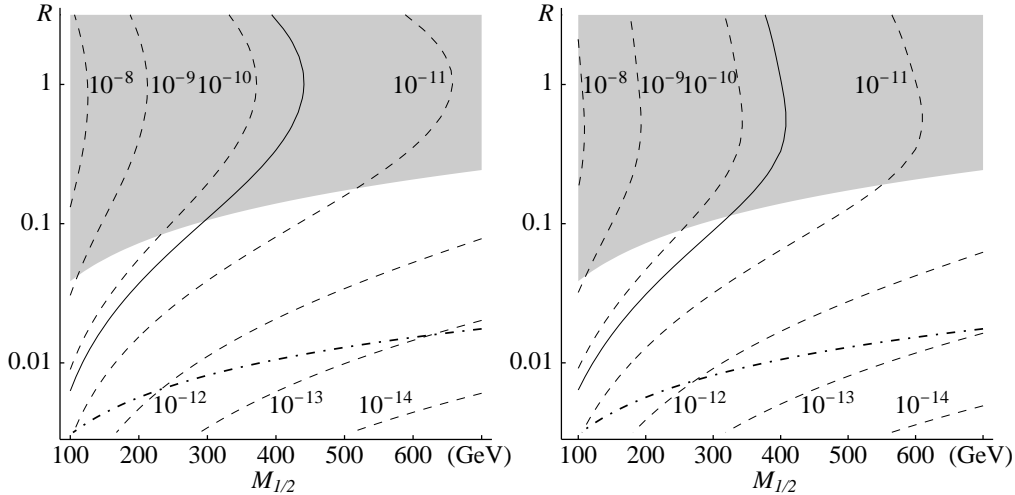}}
\caption{The branching ratio for $\mu\to e\gamma$ with the direct
contribution from the $F$-term of the dilaton ($m_0^2=M^2_{1/2}$ 
at the GUT scale). The anomalous $U(1)$ charges and the meanings of 
the lines are the same as in Fig.~\ref{fig_br_A}.}
\label{fig_br_dilaton}
\end{figure}

Figure~\ref{fig_br_dilaton} exhibits the branching ratio in the case that the
$F$-term of the dilaton also contributes to the scalar fermion  masses
directly, as $m_0= M_{1/2}$ at the GUT scale. Even if $R=0.1$, the
gaugino mass $M_{1/2}>300$GeV is allowed. However, we cannot predict
the lowest value of the branching ratio, because the upper limit of the
contribution from the dilaton is unknown in this framework. 

\section{Discussion and Summary}
There is a problem in the situation that the gaugino mass dominates
sfermion masses ($m_0/M_{1/2} \ll 1$) at the GUT scale, as discussed in
Ref.~\cite{UFB}. This is that the effective potential is unbounded from
below (UFB) and that the standard vacuum in which the electroweak
symmetry breaking occurs in a right manner is metastable. In our
scenario the ratio of sfermion masses to the gaugino mass $m_0/M_{1/2}$
is of the order of $0.3$ for $R\sim 0.1$. According to the analysis in
Ref.~\cite{UFB}, this ratio means that our scenario is located just
around the boundary of the UFB constraint and that the constraint may be 
satisfied. Even if our scenario does not satisfy it, there is a
possibility that the standard vacuum is more stable than the
universe. Moreover, if there is another flavor independent contribution
to the sfermion masses, such as from GUT interactions or the dilaton, it
can be free from this problem since $m_0/M_{1/2} \sim 1$.

In this paper, we have studied the case that the dilaton is stabilized
by the deformation of the K\"ahler potential for the dilaton and have
pointed out that the order of the ratio of the $F$-term to the $D$-term 
contributions is generally determined. We estimated FCNC processes from
this ratio in this scenario, and we showed that in particular the $\mu
\to e \gamma$ process can be around the present experimental upper
bound. The analysis in the LANL experiment is expected to make the
experimental bound to the branching ratio for $\mu\to e \gamma$ 
lower to $5\times 10^{-12}$. Moreover, in near future it 
can be expected that the process is accessible to the branching ratio 
of the order of $10^{-14}$ \cite{Kuno}. 
Therefore this process should be observable in the near future if the
scenario discussed here is true.

\section*{Acknowledgements}
We would like to thank H.~Nakano, M.~Nojiri and M.~Yamaguchi for
valuable discussions and comments. N. M. also thanks D. Wright for
reading this manuscript and useful comments. The work of N.~M. is
supported in part by a Grant-in-Aid for Scientific Research from the
Ministry of Education, Science, Sports and Culture.

\appendix
\renewcommand{\theequation}{A.\arabic{equation}}

\section*{Appendix}

In this appendix we briefly give the calculation of the ratio $r$ in
Eq.~(\ref{general_r}) for the generalized superpotential
(\ref{general_W}),
\begin{eqnarray}
W&= &C^{\bar \imath_1, \cdots , \bar \imath_n}_{i_1, \cdots , i_n}
  \left(\frac{Q^{i_1}\bar Q_{\bar \imath_1}}{\Mp^2}\right)\cdots
  \left(\frac{Q^{i_n}\bar Q_{\bar \imath_n}}{\Mp^2}\right)
  \left(\frac{\Phi}{\Mp}\right)^{q_{sum}}\Mp^3,
\end{eqnarray}
where
\begin{equation}
q_{sum}= \sum_{p= 1}^{n} (q_{i_p}+\bar q_{{\bar \imath}_p}).
\end{equation}
Below, the dynamical scale $\Lambda$ of the $SU(N_c)$ gauge theory, the
effective superpotential can be written in terms of meson fields $M^i_{\bar
\jmath}\equiv Q^i\bar Q_{\bar \jmath}$ as,
\begin{equation}
W= C^{\bar \imath_1, \cdots , \bar \imath_n}_{i_1, \cdots , i_n}
  \left(\frac{M^{i_1}_{\bar \imath_1}}{\Mp^2}\right)\cdots
  \left(\frac{M^{i_n}_{\bar \imath_n}}{\Mp^2}\right)
  \left(\frac{\Phi}{\Mp}\right)^{q_{sum}}\Mp^3
  +(N_c-N_f)\left(\frac{\Lambda^{3N_c-N_f}}{\det
       M}\right)^{\frac{1}{N_c-N_f}}.
\end{equation}
On the other hand, since the K\"ahler potential for the dilaton field
$S$ must be a function of $S+S^\dagger-\delta_{GS} V^A$, because of an
anomalous $U(1)$ gauge invariance, the total K\"ahler potential is
\begin{equation}
K_{tot}= Q^iQ^\dagger_ie^{2q_iV^A}
  +\bar Q^{\dagger \bar \imath}\bar Q_{\bar \imath}
      e^{2\bar q_{\bar \imath}V^A}
  +\Phi^\dagger \Phi e^{-2V^A}
  +K(S+S^\dagger-\delta_{GS} V^A).
\end{equation}
Along the $SU(N_c)$ flat direction, the total K\"ahler potential can be
written as
\begin{equation}
K_{tot}= \left(\sqrt{M M^\dagger}\right){}^i_i e^{2q_iV^A}
       +\left(\sqrt{M^\dagger M}\right){}^{\bar \imath}_{\bar \imath}
      e^{2\bar q_{\bar \imath}V^A}
  +\Phi^\dagger \Phi e^{-2V^A}
  +K(S+S^\dagger-\delta_{GS} V^A).
\end{equation}
Then
\begin{eqnarray}
\int d^4\theta K_{tot}
  &= &-D_A\left\{q_i\left(\sqrt{M M^\dagger}\right){}^i_i 
             +{\bar q}_{\bar \imath}\left(\sqrt{M^\dagger M}\right)
                {}^{\bar \imath}_{\bar \imath}
             -|\Phi|^2 + \xi^2
	\right\}
     \nonumber\\
  && +F_M{}^k_{\bar k}F_M^\dagger{}^{\bar l}_l
      (K_{MM^\dagger}){}^{\bar k}_k\;{}^l_{\bar l} 
 \hspace{0.5em}+ \cdots ,
\end{eqnarray}
where
\begin{eqnarray}
(K_{MM^\dagger}){}^{\bar k}_k\;{}^l_{\bar l}
 \equiv \frac{\del^2 \left\{\Tr\left(\sqrt{M M^\dagger}\right) 
             +\Tr\left(\sqrt{M^\dagger M}\right)\right\}
          }
         {\del M^k_{\bar k} \del M^\dagger{}^{\bar l}_l},
\end{eqnarray}
and $\xi^2$ is given by Eq.~(\ref{FIterm}). The total scalar potential is
\begin{equation}
V= \frac{1}{K''}\left|\frac{\del W}{\del S}\right|^2
  +\left| \frac{\del W}{\del \Phi}\right|^2
  + \frac{\del W}{\del M^k_{\bar k}} 
    (K^{-1}_{MM^\dagger}){}_{\bar k}^k\;{}_l^{\bar l}
    \frac{\del W^*}{\del M^\dagger{}^k_{\bar k}}
  + \frac{1}{2g_A^2}|D_A|^2.
\end{equation}
Here we define the parameter $\epsilon \ll 1$ by 
\begin{equation}
  \epsilon \equiv  \left(\frac{\Lambda}{\Mp}\right)
             ^\frac{3N_c-N_f}{n N_c-(n-1)N_f},
\end{equation} 
so that we can find a local minimum as an expansion in $\epsilon$:
\begin{equation}
\VEV{\Phi^2}= \xi^2 + O(\epsilon).
\end{equation}
At the minimum, the VEVs of the meson fields are given by the solution of
the equations,
\begin{equation}
nC^{\bar \imath_1,\bar \imath_2 \cdots , \bar \imath_n}
   _{i_1,i_2, \cdots , i_n}
  \left(\frac{M^{i_2}_{\bar \imath_2}}{\Mp^2}\right)\cdots
  \left(\frac{M^{i_n}_{\bar \imath_n}}{\Mp^2}\right)
  \left(\frac{\xi}{\Mp}\right)^{q_{sum}}\Mp
-(M^{-1}){}^{\bar \imath_1}_{i_1}
  \left(\frac{\Lambda^{3N_c-N_f}}{\det M}\right)^{\frac{1}{N_c-N_f}}
= 0,
\end{equation}
and the VEVs of the auxiliary fields $D_A$ and $F_S$ can be written as
\begin{eqnarray}
-\VEV{D_A}
&=&  \epsilon^{2n} \frac{\Mp^6}{\xi^4}
    \left|\frac{1}{\det m}\right|^{\frac{2}{N_c-N_f}}(2C_N)^2
    \nonumber\\
&&\times\left\{ \frac{2C_N}{N_c-N_f+\frac{N_f}{n}}
     \left(1-\frac{2K'}{\delta_{GS}K''}\right)-1 \right\},\\
\VEV{F_S}
&=&  \epsilon^n \frac{\Mp^3}{\xi^2} (2C_N) \frac{K'}{K''}
    \left( \frac{1}{\det m^\dagger}\right)^\frac{1}{N_c-N_f},
\end{eqnarray}
where
\begin{equation}
m^i_{\bar \imath}\equiv \frac{\VEV{M^i_{\bar \imath}}}{\epsilon \Mp^2}.
\end{equation}
Then the ratio $r\equiv|\VEV{D_A}|/|\VEV{F_S}|^2$ is
\begin{eqnarray}
r=  \frac{1}{N_c-N_f+\frac{N_f}{n}}\frac{4C_N}{\delta_{GS}}
   \left(\frac{K''}{-K'}\right)
  +\left\{\frac{2C_N}{N_c-N_f+\frac{N_f}{n}}-1\right\}
   \left(\frac{K''}{-K'}\right)^2.
\end{eqnarray}
Moreover, we can get the relation among derivatives of the K\"ahler
potential using the minimization condition for the dilaton field $S$:  
\begin{equation}
\frac{K'''}{K'}= -\frac{1}{N_c-N_f+\frac{N_f}{n}}
                \frac{4C_N}{\delta_{GS}}\frac{K''}{K'}
               \left(1-\frac{\delta_{GS}K''}{2K'}\right)^2
              +\frac{\delta_{GS}}{2}\left(\frac{K''}{K'}\right)^3.
\label{exact_ratio}
\end{equation}
%


\end{document}